\documentclass[runningheads]{llncs}
\usepackage{graphicx}

\usepackage{booktabs}
\usepackage{url}
\usepackage[protrusion=true, expansion=true, final]{microtype} 
\def\shortname{\textsc{NLPExplorer}}

\begin{document}
\title{\shortname: Exploring the Universe of NLP Papers}
%

\newcommand{\repeatthanks}{\textsuperscript{\thefootnote}}

\author{Monarch Parmar\thanks{Alphabetically ordered with equal contribution} \and
Naman Jain\repeatthanks \and
Pranjali Jain\repeatthanks \and
P Jayakrishna Sahit\repeatthanks \and
Soham Pachpande\repeatthanks \and
Shruti Singh \and
Mayank Singh
}

\authorrunning{Parmar et al.} 
%
\institute{Indian Institute of Technology Gandhinagar, Gandhinagar, Gujarat, India  \\
\email{singh.mayank@iitgn.ac.in}
}
\maketitle              
\begin{abstract}
Understanding the current research trends, problems, and their innovative solutions remains a bottleneck due to the ever-increasing volume of scientific articles.  In this paper, we propose~\shortname, a completely automatic portal for indexing, searching, and visualizing Natural Language Processing (NLP) research volume. \shortname~presents interesting insights from papers, authors, venues, and topics. In contrast to previous topic modelling based approaches, we manually curate five course-grained non-exclusive topical categories namely \textit{Linguistic Target} (Syntax, Discourse, etc.), \textit{Tasks} (Tagging, Summarization, etc.), \textit{Approaches} (unsupervised, supervised, etc.), \textit{Languages} (English, Chinese, etc.) and \textit{Dataset types} (news, clinical notes, etc.). 
Some of the novel features include a list of young popular authors, popular URLs and datasets, list of topically diverse papers and recent popular papers. Also, it provides temporal statistics such as yearwise popularity of topics, datasets, and seminal papers.  
To facilitate future research and system development, we make all the processed dataset accessible through API calls. The current system is available at \url{http://lingo.iitgn.ac.in:5001/}.

\end{abstract}

\section{Introduction}
Effective scientific literature understanding plays a critical role towards the research community's common goal of \textit{``March for Science''}. However, the yearly generated research volume shows an upward trend with an estimated increase of 8-9\% per year. This results in information overload, impacting the literature review process and often, leads to \textit{`reinventing the wheel'} syndrome. This negatively affects the efficiency of scientific progress on the knowledge frontier. 

In the past, significant efforts have been made to curate  peer-reviewed open access scientific information. In particular, the field of Natural Language Processing ($NLP$) witnessed the development of \textit{ACL Anthology} since the year 2001, which curates papers (in PDF format) from more than 70 NLP venues including popular conferences like ACL, NAACL, EMNLP, etc. In the year 2008, Bird et al.~\cite{bird2008acl} released the \textit{ACL Anthology Reference Corpus (ACL ARC)}, consisting of OCRed extracted text and metadata of PDF articles. In the year 2009, Radev et al.~\cite{radev2009acl} developed \textit{ACL Anthology Network (AAN)}  by manually constructing paper citation, author citation and author collaboration network, along with other interesting statistics and citation summaries of the articles. The recently updated AAN system~\cite{radev2013acl} indexes articles published till 2014. Another initiative, ACL Anthology SearchBench~\cite{schafer2011acl}, provides bibliographic metadata filtering and full-text structured semantic search in ACL Anthology. CL Scholar~\cite{singh2018cl} periodically crawls ACL Anthology and constructs a computational linguistic knowledge graph. It supports natural language queries and entity specific queries about the authors, venue, and paper. Almost all systems described above are either in dormant condition~\cite{bird2008acl,radev2009acl,radev2013acl,schafer2011acl} or not available~\cite{singh2018cl}.

In this paper, we discuss the development of \shortname. \shortname~provides paper, author, venue and topic-specific temporal as well as aggregated statistics. For the first time, we also showcase, the usage of URLs over the years, popular top-level and sub-domains, survey papers, new authors, etc. The system also presents a visualization of the timeline for the first occurrences of sub-topics. 

Our main contributions are as follows:
(i) We periodically download, preprocess, and index the ACL Anthology dataset consisting of more than 55 thousand full-text PDF articles. (ii) We invest extensive effort in automatic curation, structured information extraction, cleaning, indexing, and other related preprocessing steps. (iii) We classify papers into a first-of-its-kind detailed list of NLP topics and subtopics. (iv) The proposed system presents content as well as bibliographic statistics along with basic keyword-based search facilities. (v) We deploy the current system in Google Cloud with an API-based retrieval facility.

\section{Dataset}
We leverage the ACL Anthology dataset~\cite{aclanthology} that hosts articles dedicated to Computational Linguistics and Natural Language Processing. Each article is present as a PDF file along with the metadata namely author list, venue name, year of publication and a unique eight-character identifier. Overall, the dataset has 55,565 papers, 39,555 unique authors, 78 venues and 723,976 citations. 

\section{Architecture}
\label{sec:architecture}
\begin{figure*}[t]
    \makebox[\textwidth][c]{\includegraphics[width=1.0\textwidth]{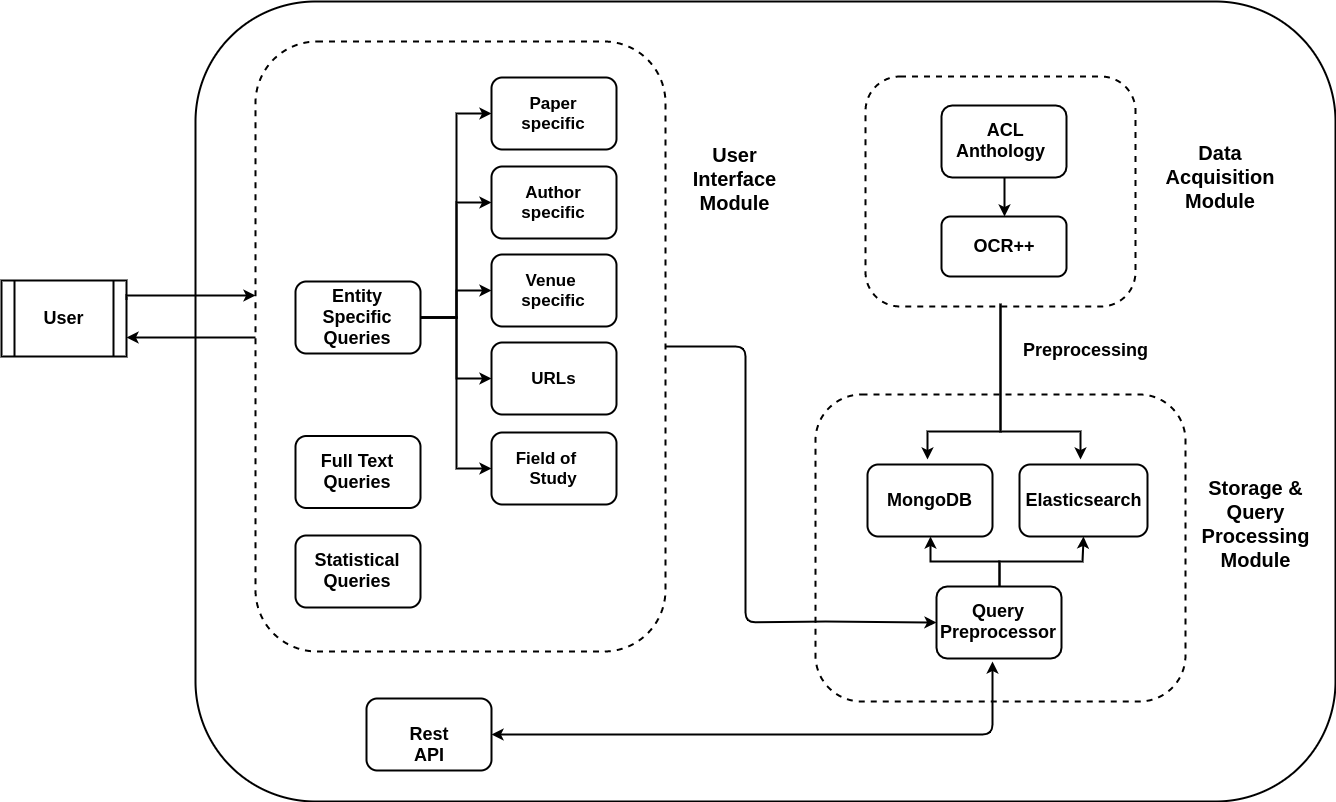}}
    \caption{Architecture of~\shortname. Arrows represent data flow.}
    \label{fig:architecture}
\end{figure*}
The architecture of~\shortname~is composed of three interdependent modules, (i) data acquisition, (ii) storage and query processing, and (iii) user interface module. Figure~\ref{fig:architecture} presents the detailed description of the architecture.  The \textit{data acquisition} module periodically curates newly added semi-structured information such as metadata and the corresponding PDF articles from ACL Anthology. It leverages OCR++~\cite{singh2016ocr++} tool to extract structural and bibliographical information from each of the PDF articles, and then passes the metadata, structural and bibliographical information to the storage and query processing module. The \textit{storage and query processing} module indexes the periodical updates into MongoDB~\cite{mongodb} database and Elasticsearch~\cite{elasticsearch} engine.  MongoDB database handles the basic storage and retrieval, whereas Elasticsearch supports full-text based query search and ranking. The third module, \textit{the user interface}, fetches processed search results from the storage and query processing module. The user interface is designed using Python's Flask library, and JavaScript libraries Plotly~\cite{plotlyjs} and Timeline~\cite{timelinejs} are used to render statistics and graphical components of the interface. 

The current system is deployed at Google cloud~\cite{googlecloud}. The infrastructure consists of a 4 CPU - 7.5GB RAM, Linux VM instance that can be extended on demand. The system supports REST API requests served using MongoDB database.

\section{Current Features}
We categorize the wide range of functionalities of \shortname~as follows:
\begin{itemize}
    \item \textbf{Entity-specific \& Full-Text search:}
    \shortname~supports basic keyword based search leveraging the metadata and the full-text of articles. The search can be chosen to output results to any of the following domains - authors, papers, venues, URLs, and field of study. We also extend full-text search of articles in order to visualise n-gram trends over the period of time.
    \item \textbf{Paper statistics:}
    We provide standard paper related statistics such as the publication year and venue, author information, citation distribution over the years and the link to the corresponding PDF article. Additionally, we provide interesting insights like similar papers, topical distribution and mentioned URLs. We also provide statistics such as the list of popular recent papers, popular survey papers, seminal papers, papers with diverse topics, and publication count of last five years.
    \item \textbf{Author statistics:}
    We provide author statistics such as publication and citation distribution over the years, topical distribution of papers and venue preference. Additionally, we provide the list of popular recent authors, popular authors in the lifetime of ACL Anthology, authors with top publication counts, recent authors with high publication counts, and highly diverse authors.
    \item \textbf{Field of study statistics:}
    We curate a comprehensive list of topics categorized into five broad categories, (i) \textit{Linguistic Targets} (Syntax, Discourse, Pragmatics, etc.), (ii) \textit{Tasks} (Tagging, Summarization, Chunking, etc.), (iii) \textit{Approaches} (unsupervised, supervised, etc.), (iv) \textit{Languages} (English, Chinese, Hindi, etc.), and (v) \textit{Dataset Types} (news, clinical notes, etc.). ~\shortname~provides interesting insights such as the temporal distribution of papers and authors in each topic and subtopics, distribution of papers in each conference, and a timeline to visualize the introduction of new topics. We make the list of topics and the processed data of the system available at the systems' webpage~\cite{nlpexp}.
    \item \textbf{Venue statistics:}
    Venue-related statistics include temporal distribution of publications and citations, topical distribution, and the list of papers in a year. We also include insights such as the top NLP venues citing and cited by the candidate venue, popular authors publishing in the candidate venue, and the shift in topical distribution over the years.
    \item \textbf{URL statistics:}
    We analyse URLs reported in the research papers. The URL-related statistics include top URLs in different categories such as universities, digital libraries, datasets and research groups, alongwith the analysis of top-level domains (TLDs) and corresponding sub-domains. Additionally, ~\shortname~provides year-wise usage distribution, total usage, and the list of top-most subdomains and associated papers.
    
\end{itemize}

\section{Future Extensions and Conclusion}

The current system provides a basic functionality for knowledge exploration in NLP domain. In future, we plan to incorporate advanced set of functionalities leveraging natural language understanding of research papers. Some of the main proposals include natural language query retrieval, intelligent ranking by leveraging citation sentiments and discourse-level citation information, visualization of topical flow from cited papers to the main text of citing papers, automatic generation of leaderboard for NLP tasks, visualization of author collaboration networks, paper citation networks, venue interaction networks, etc.

In this paper, we present an end-to-end automated system that periodically mines the ACL Anthology and serves as a tool to aid researchers in knowledge exploration and discovery. The goal of \shortname~is to serve as a retrieval engine for research papers, as well as a tool to assist researchers in knowledge discovery by helping them to better understand the problem domain, the top researchers in the field of study, and the latest research in the domain. Even though, the current system supports NLP research domain, we claim that similar systems can be built for any domain given the availability of full text articles and basic metadata. 

\bibliographystyle{splncs04}
\bibliography{main}

\end{document}